\documentclass[
    ,final            
  ]
  {aipproc}

\layoutstyle{6x9}

\begin{document}

\title{A ``Crib Sheet'' for Supernova Events}


\author{David Arnett}{
  address={Steward Observatory, University of Arizona, Tucson AZ 85721, USA}
  ,altaddress={ICRA, Rome, Italy}
}



\begin{abstract}
This paper summarizes our theoretical understanding of supernova
events in a ``back of the envelope'' way. It is intended to aid in
the recognition and understanding of those events which are not
``standard'', and which may provide the most insight.
\end{abstract}

\maketitle


\section{Introduction}
This material is abstracted from previous work by the author;
see \citet{arnett1996,arnett1977} for a more detailed discussion and 
references to earlier work. It is hoped that by using the template analysis 
given below, it will be easier to recognize non-standard events.

\section{A ``Standard'' Supernova}
We define a standard supernova event (SN) and show how to scale 
to nonstandard cases. Our standard SN is assumed to have
\begin{itemize}
\item a luminosity of $L = 10^{10}\ \rm L_\odot$, 
\item an effective temperature of $T \approx 10^4\ \rm K$, and 
\item will stay bright for a time of a few weeks ($2 \times 10^6 \rm\ sec$).
\end{itemize}
This implies several interesting things: 
\begin{itemize}
\item Using 
$L/L_\odot = (R/R_\odot)^2(T_e/(T_e)_\odot)^4$,
the radius is  $R \approx 1.2 \times 10^{15}\ \rm cm$. This is 
about 20 times larger than the radii of the largest supergiants. 
\item To move this distance
in $2 \times 10^6 \rm\ s$ implies an \emph{average} velocity 
$v \approx 6 \times 10^8 \rm \ cm/s$. The photospheric sound speed
is only $\sqrt{1.5 R_{gas}T} \approx 10^6 \rm \ cm/s$, so the flow is
highly supersonic. 
\item The energy radiated as visible light is 
$Lt \approx 2 \times 10^{49}\ \rm ergs$. Suppose for example, this is 
fully supplied by radioactive decay of $\rm Ni^{56}$. At about $ 2 \ \rm MeV$
of plasma heating per decay, this implies more than $0.1\ \rm M_\odot$ of
that nucleus be freshly produced in the SN\footnote{Fast production
involves a time scale short compared to electron capture time in the
dense $\rho > 10^5 \ \rm g/cm^3$ plasma, which is itself much shorter
than the half-life of 5.9 days. Further, the time from synthesis to maximum
light allows significant  $\rm Ni^{56}$ decay (see \cite{arnett1996}).}.
\end{itemize}

\section{Radiative Diffusion Time Scale}
The time scale for radiative diffusion of energy from a sphere of radius 
$R$ is $\tau_d = 3R^2/\pi^2 \lambda c$, where the mean-free-path is 
$\lambda = 1/ \rho \kappa$, with $\rho$ the mass density and $\kappa$ 
the Rosseland mean opacity. We set this equal to the expansion time scale,
$\tau_e = R/v$, where $v$ is the velocity of expansion, and of order the
average velocity we derived aboved. The average density is 
$\rho= 3M/4\pi R^3$, so we obtain a time scale 
\begin{equation}
\tau \approx 3 \times 10^6 \ {\rm s} \left [ 
\left ({\kappa \over 0.2\ {\rm cm^2/g}} \right )
\left ( {M \over M_\odot}\right )
\left( { 10^8 {\rm \ cm/s} \over v} \right ) \right ] ^{1 \over 2}, 
\label{tdiff}
\end{equation}
which is essentially the time to maximum light for an expanding, diffusing
sphere (see \S~13.3 in \cite{arnett1996}). 
If the opacity $\kappa$ is near the Thompson scattering value 
(0.4 for pure H and 0.2 for pure $\rm He^4$), that factor is of order unity.
Taking $\tau \approx 2 \times 10^6 \ \rm s$
and $v \approx 6 \times 10^8 \rm \ cm/s $,
we get $ M \approx 2.6 M_\odot$ or so. These expressions are approximate;
see \cite{arnett1996} or recent simulations for more precise values. 
Nevertheless we see that the mass of the exploding object is star-sized,
yet less than the most massive stars (probably due to mass loss).

Using this mass and radius, the average density is 
$\rho = 3M/4\pi R^3 \approx 10^{-12} \rm \ g/cm^3$, so that for 
$\kappa = 0.2$ the mean-free-path  is 
$\lambda =1/\rho \kappa \approx 5 \times 10^{12} \rm\ cm$.
The object is therefore 240 mean-free-paths deep. This justifies using the
diffusion time, but also indicates that supernovae are a transition between
optically-thick stars and optically-thin nebulae.

\section{Adiabatic Cooling and Kinetic Energy}
Taking this mass and the average velocity, the kinetic energy is 
$K \approx 10^{51} \ \rm erg$, or one bethe, 
so that the radiated photon energy
is only about 0.02 of this. Adiabatic cooling converts internal energy into
kinetic energy, so that for this radiation-dominated gas, this corresponds to
an initial radius smaller by the same factor, 0.02, which gives 
$R(0) \approx 2.4 \times 10^{13}\ \rm cm$. This is in the range of radii 
of red giants. Shock heating can power the light curve only if the progenitor
has a radius this large; otherwise the shock energy is degraded into
kinetic energy of expansion. To the extent that the light curve is dominated
by shock energy, a large radius means a bright, slow event, and a small 
radius means a dim, fast event. 

Mass loss by single massive stars and by binary interactions is
an important parameter for interpreting supernovae behavior 
\citet{arnett1977}.
As Eq.~\ref{tdiff} indicates, a larger mass gives a slower light curve by
increasing the diffusion time\footnote{Increased opacity gives a similar
effect.}. 

Unless there is matter to collide with at
radii $R > 10^{15}\ \rm cm$, the kinetic energy of expansion is unable 
to provide radiation. How much mass loss would be required to provide
enough circumstellar matter to affect the
light curve? We consider one interesting case: suppose that the 
progenitor is a red supergiant losing mass with a velocity 
$w \approx 10^6 \rm \ cm/s$. The uniform mass loss rate for the
progenitor would be $dM_p /d t = 4 \pi r^2 \rho(r) w$. The SN shock would
collide with this pre-existing mass distribution, raising the energy of
this material at a rate
$ L_K = (v^2/2) dM_{shock}/dt$, where the rate of sweeping up of
mass is $dM_{shock} = 4 \pi r^2 \rho(r) v$. Substituting $r^2 \rho(r)$
we have from the progenitor mass loss,
$L_K = (dM_p/dt) v^3/2w $, or
\begin{equation}
L_K/10^{43}{\rm\ erg/s} \approx 7.2 \times 10^2 \left ( {dM_p \over dt}/ 
{\rm M_\odot {yr}^{-1}} \right ).
\end{equation}
A mass loss rate of $10^{-4}\rm\ M_\odot\ {yr}^{-1}$ 
from the progenitor in a time
$t > R/w \approx 40 \rm\ years$ prior to the explosion
implies a significant source of energy. 
Does it make a visible display? Where does this energy go?
The collision occurs in optically-thin plasma, under non-equilibrium 
conditions;
we are in a ``nebular'' regime. The collision generates turbulence, and that
generates magnetic field. Being optically thin, significant radio and
x-ray emission occurs.
Cosmic-ray acceleration should result 
(\citet{ginsburg1964}). See \citet{fransson1996} for a detailed analysis of the
well-observed SN1993J.

\section{Radioactivity}
Radioactive decay energy cannot be adiabatically cooled until the decay
actually happens, so radioactive energy is automatically ``saved'' until a 
time corresponding to the mean lifetime of the decay. Each 
radioactive species adds an exponential decay component to 
the light curve. For this to be significant, a large amount of mass must 
be synthesized into a suitable nucleus.
$\rm Ni^{56}$ is exceptional in this regard: it is doubly-magic, and the 
most bound nucleus having equal numbers of neutron and protons $Z=N$, 
so it is easy to synthesize by explosions from fuels with $Z=N$. The ashes
of all hydrostatic burning stages up to oxygen consumption have $Z=N$ to a
suffienctly good approximation to allow profuse production of $\rm Ni^{56}$.
It is the dominant radioactivity found so far.  $\rm Ni^{56}$ decays to
 $\rm Co^{56}$ by electron-capture with a 5.9 day halflife. The Co
decays by both electron-capture and by positron-emission to  $\rm Fe^{56}$
with a 77.3 day halflife. The characteristic light curves of 
type~I supernovae reflect this double decay, with their 
``decline from peak'' and ``tail'' \footnote{The rise to peak is shaped 
by radiative diffusion \citet{arnett1996}}.
More massive ejecta have longer radiative diffusion times, and can smear out
the $\rm Ni^{56}$ peak, leaving only the $\rm Co^{56}$ tail.
The stable product, $\rm Fe^{56}$, is the sixth most abundant nucleus in the
solar-system abundance distribution, and almost all is believed to be formed
in supernova explosions.

There are a number of other important radioactive nuclei, such as
$\rm Ti^{44}$ and $\rm Al^{26}$, which act as tracers of nucleosynthesis
but none have yet been found which dominate the light curve so much as the 
$\rm Ni^{56}$-$\rm Co^{56}$-$\rm Fe^{56}$ chain.

\section{Limits to $\rm\ Ni^{56}$ Production}

Because of its profound affect on light curves, it is interesting to ask
if it is possible for such explosions to produce little $\rm Ni^{56}$. 
Do all of these explosions have radioactive heating?
There are at least two obvious cases in which the $\rm Ni^{56}$ production is 
limited:
\begin{description}
\item[Density gradient:] if the progenitor has a very steep density gradient
around the core, there will be be little appropriate fuel.
The mass of $\rm Z=N$ matter, at right radius to be shocked hot enough
to make $\rm Ni^{56}$, will be too small for much production. 
This happens for stars with cores very close to the Chandrasekhar mass 
(e.g., $10 \rm\ M_\odot$).
\item[Accretion:]  $\rm Ni^{56}$ is made but collides with a massive mantle,
is slowed, and accreted onto a newly formed neutron star or black hole. This
happens for fairly massive cores (e.g., stars $35 \rm\ M_\odot$ on the main
sequence, having helium cores of $10 \rm\ M_\odot$).
\end{description}

\section{Sources of Light Curve Energy}
Some possible sources of internal energy to supply supernova light curves:
\begin{description}
\item[Shock Breakout]
 The emergence of the supernova shock from the progenitor star.
Roughly speaking, this occurs when the photon diffusion time equals
the time for the shock to reach the stellar surface. 
\item[Diffusion of Shock Energy]
 The leakage of shock-generated heat by radiative diffusion, after
the shock has gone beyond the stellar surface (SNII, e.g.).
\item[Diffusion of Heat from Radioactivity]
The heating by radioactive decay counters adiabatic cooling, and
powers the light curve (SNIabc, SN1987A, e.g.)
\item[Heating by New Neutron Star]
MHD driven by rotational energy (Crab Nebula pulsar), or late accretion
(Not recognized?)
\item[Heating by Accretion onto New Black Hole] 
(Not recognized?)
\item[Heating related to Binary Companion]
(Not recognized?)
\end{description}
Notice that several of these possibilities have not been recognized,
and hence might be interesting to look for.

\section{Prediction of Unobserved Light Curves}

With the rapid advance of astronomical technology, it is interesting to
ask if there are types of events that we have not yet seen, but might.
In this regard we note that what are now called supernova of Type~Ib and Ic,
were suggested to exist before they were recognized; they were the 
``bare cores'' discussed in \citet{arnett1977}. \citet{elias1985} suggested
that their observational data on 11 Type~I supernovae fell into two 
classifications, Ia and Ib. The brighter Ia's correspond to exploding
white dwarf stars, for which analytic solutions for the light curves
were available \citet{arnett1982}. The dimmer Ib's correspond to
core collapse events which produce $\rm Ni^{56}$, which \citet{colgatemckee}
first introduced to explain all types of supernovae.
\citet{arnett1977} had suggested that both types should be found, and that
the core collapse events would be dimmer because they would tend to
make less  $\rm Ni^{56}$. \citet{filippenko}
found a further class, Ic's, which unlike Ib's, are He poor; apparently
they are bare cores that lost He as well as H. \citet{wheeler} give a
review of the observational situation at the time.

\subsection{A Table of Some Possibilities}

\begin{table}
\begin{tabular}{lrrrrrr}
\hline
   \tablehead{1}{r}{b}{$R(0)$} &
   \tablehead{1}{r}{b}{$M_{ej}$} &
   \tablehead{1}{r}{b}{Radioactivity} &
   \tablehead{1}{r}{b}{Brightness} &
   \tablehead{1}{r}{b}{Speed}  &
   \tablehead{1}{r}{b}{Example} \\
\hline
small & small  & large & bright & fast & SNIa \\
large & large  & some  & medium & slow & SNIIP \\
large & small & some  & medium & slow & SNIIL \\
small & small  & some  & medium & fast & SNIb,c \tablenote{Predicted in \citet{arnett1977}}\\
small & large  & some  & medium & slow & SN1987A \\
large & small  & some  & medium & fast & SN1993J \\
large & small  & no    & medium & fast\tablenote{shock} & \\
small & small  & no    & dim    & fast & \\
large & large  & no    & dim    & slow & \\

\hline
\end{tabular}
\caption{SN Parameters and Light Curves}
\label{tab:a}
\end{table}

Table~\ref{tab:a} sketches the sorts of exploding supernovae
we might expect to find; other entries may need to be added.
The parameters varied here, besides initial mass, are initial
radius $R(0)$, mass ejected by explosion $M_{ej}$, and presence
of radioactivity. Even this limited set covers much of the
observed data set, at least in a crude way. But is the limited
set of paramaters sufficient for a more complete and accurate
data set?

\section{Some Final Comments}

Some key points: 
\begin{itemize}
\item It is not theoretically demanded that all
``supernova'' explosions have radioactive ejecta and/or large
initial radii, the two features that make them bright.
Therefore, explosions of small initial radii and little
radioacitity in their ejecta would be under-represented
in the present data set. Such events would be dim and fast.

\item None of these entries are related to binary interactions
except is a fairly passive way of slow mass transfer for SNIa's. 
Are there merger supernovae (for example two white dwarfs or
a white dwarf and a neutron star or black hole)?

\item Mass loss, either by winds from single
stars or by binary interactions, is an importent parameter
for the observed explosion; mass loss is not well understood
from a predictive basis yet. 

\item Rotation and magnetic fields
provide vector fields which interact and which we are just
beginning to learn to simulate plausibly; they are expected
to be important aspects of the understanding of the formation
of relativistic jets and GRB's, as well as the general issue of
angular momentum transport in stars.

\item Fully 3D simulations of oxygen and silicon burning are in their
infancy (\citet{meakin2007,meakin2006}), 
and already there are indications of complexity we
have not anticipated. For example, 2D simulations of C, Ne, O, and Si shells
indicate that there are
interactions between burning shells which are mediated by
waves (p-, g-, and mixed-modes), and none of this has yet been put
into evolutionary models.

\item Extrapolation to earlier cosmological epochs is made more
uncertain by a lack of fundamental understanding of how magnetic fields,
rotation, binary formation, and mass loss change with metalicity,
yet interpretation of some of the new supernova observations may be 
affected by these issues.
\end{itemize}

Despite these theoretical challenges, the prospect for observation
of previously inaccessible phenomena is excellent. For supernova
studies, the exploration of a broad time domain and greater
completeness is exciting. It would be both a surprise and
a disappointment if some really new phenomena, either
unrecognized or unexpected, do not appear soon!


\begin{theacknowledgments}
This work was supported in part by  NSF Grant 0708871 and NASA Grant
NNX08AH19G at the University of Arizona. The kind hospitality of 
Prof. Sandip K. Chakrabarti and the Bose Center for Fundamental Research 
is gratefully acknowledged.

\end{theacknowledgments}



\bibliographystyle{aipprocl} 




\end{document}